\documentstyle[eqsecnum,floats,prd,aps,epsf]{revtex}

\begin{document}

\twocolumn[\hsize\textwidth\columnwidth\hsize\csname@twocolumnfalse\endcsname

\title{A numerical testbed for singularity excision 
	in moving black hole spacetimes}

\author{Hwei-Jang Yo$^{1,2}$, Thomas W. Baumgarte$^{3,1}$,
        and Stuart L. Shapiro$^{1,4}$}

\address{$^1$Department of Physics,
University of Illinois at Urbana-Champaign, Urbana, IL 61801\\
$^2$Institute of Astronomy and Astrophysics, Academia Sinica,
Taipei 115, Taiwan, Republic of China\\
$^3$Department of Physics and Astronomy, Bowdoin College, Brunswick, ME 04011\\
$^4$Department of Astronomy \& NCSA, University of Illinois at
Urbana-Champaign, Urbana, IL 61801}

\maketitle

\begin{abstract}
We evolve a scalar field in a fixed Kerr-Schild background geometry to
test simple $(3+1)$-dimensional algorithms for singularity excision.
We compare both centered and upwind schemes for handling the shift
(advection) terms, as well as different approaches for implementing
the excision boundary conditions, for both static and boosted black
holes. By first determining the scalar field evolution in a static
frame with a $(1+1)$-dimensional code, we obtain the solution to
very high precision.  This solution then provides a useful testbed for
simulations in full $(3+1)$ dimensions.  We show that some algorithms
which are stable for non-boosted black holes become unstable when the
boost velocity becomes high.
\end{abstract}
\vskip2pc]

%===================================
\section{Introduction}
%===================================
\label{intro}
%===================================

The long-term numerical evolution of black holes is one of the most
important and challenging problems in numerical relativity.
Simultaneously, it is a problem for which a solution is very urgently
needed; binary black holes are among the most promising sources for
the gravitational wave laser interferometers currently under
development, including LIGO, VIRGO, GEO, TAMA and LISA, and
theoretically predicted gravitational wave templates are crucial
for the identification and interpretation of possible signals.

Numerical difficulties arise from the complexity of Einstein's
equations and the existence of a singularity inside the black hole
(BH). Numerical simulations based on the traditional
Arnowitt-Deser-Misner (ADM) decomposition in $(3+1)$ dimensions, for
example, often develop instabilities \cite{bonc2,baut}.  The gauge
(coordinate) freedom inherent to general relativity constitutes a
further complication.  Singularity avoiding slicings
\cite{smal,eard,barj} can follow evolutions involving black holes only
for a limited time, since the stretching of time slices typically
causes simulations to crash on time scales far shorter than the time
required for a binary BH orbital period.

Recently, there have been several very promising numerical
breakthroughs. Stable formulations of Einstein's equations using a
conformal-tracefree decomposition have been developed by Shibata and
Nakamura \cite{shim} and by Baumgarte and Shapiro \cite{baut}. The
so-called BSSN formulation has shown remarkable stability properties
when compared to the ADM formulation in a wide range of numerical
simulations \cite{baut2,shim2,shim3,shim4,alcm4,alcm5}. Also, the
development of ``singularity excision'' techniques
\cite{thoj2,seie,thoj3,annp,schm2}, excising the singularity from the
computational domain, may allow for long term binary BH evolutions
(see \cite{bras} for preliminary results).  There have also been
proposals \cite{matr,bisn,marp} for alternative families of binary
initial data based on the Kerr-Schild form of the Schwarzschild (or
Kerr) metric to represent each of the BHs. The use of Kerr-Schild
coordinates is desirable for the numerical evolution of BHs and
suitable for applying singularity excision techniques because the
coordinates smoothly penetrate the horizon of the holes.  Moreover,
they provide a natural framework for constructing initial data without
assuming conformal flatness, essential for representing non-distorted 
Kerr Black holes.

Typically, the numerical grid extends into the black hole in
singularity excision applications.  Moreover, the shift vector
sometimes becomes larger than unity, so that some finite difference
stencils become ``acausal'' and therefore unstable.  To avoid this
problem, ``causal differencing'' \cite{seie,annp} or ``causal
reconnection'' schemes \cite{alcm} have been suggested (see also
\cite{schm2,gunc,schm,annp2,brub,coog,lehl}).  Unfortunately, these
schemes are fairly involved and complicated.

Alcubierre {\it et al} \cite{alcm2,alcm3} recently proposed a simple
BH excision algorithm for a non-boosted distorted BH evolution which
avoids the complications of causal differencing.  Their algorithm is
based on the following simplifications: (a) excise a region adapted to
Cartesian coordinates, (b) use a simple inner boundary condition at
the boundary of the excision zone, (c) use centered differences in all
terms except the advection terms on the shift, where upwind
differencing along the shift direction is used.  Their algorithm is
quite successful and allows for accurate and stable evolution of
non-boosted distorted BHs for hundreds of dynamical times.

Here we devise an experiment to test some simple singularity excision
algorithms for evolving dynamical fields in numerical 3D black hole
spacetimes.  Our goal is to identify a numerical scheme which is
simple and stable like the one presented in
\cite{alcm2} for non-boosted BHs but can also handle moving BHs. Our
experiment consists of tracking the propagation of a scalar field in
the fixed background spacetime of a Kerr-Schild BH. By first solving
for the evolution in a static frame by means of a 1D code, we obtain
the solution to arbitrary accuracy. We then perform simulations in
full 3D for both stationary and boosted Kerr-Schild BHs, using our 1D
``exact'' results for detailed comparison. We use this testbed to
compare schemes which handle advection terms by centered versus upwind
differencing. We also run our experiment for different implementations
of the excision boundary conditions.  Our aim is to find numerical
stencils which we can later adapt to a BSSN scheme for evolving
dynamical spacetimes with moving (e.g., binary) BHs.  While the
proving ground we construct here employs a fixed background geometry,
we feel that it furnishes necessary, if not sufficient, conditions
that must be met for a field evolution scheme in any numerical
spacetime containing BHs. We provide this report to convey the utility
of this testbed as a quick diagnostic of alternate differencing
stencils and with the hope that it might prove helpful for other 3D
code builders.

Our results show that, in general, an upwind scheme is more stable
than a centered scheme.  This is consistent with the results of
\cite{alcm2}.  We also find that a higher resolution is needed for an
upwind than for a centered scheme to achieve a desired accuracy, due
to the diffusive character of an upwind scheme.  We find that the
numerical implementation used in \cite{alcm2} is stable in the
non-boosted case but unstable in the boosted case.  However, the
stability can be restored by using an alternative excision boundary
condition which is stable in both the non-boosted and boosted
cases. In all the cases studied, the use of a third-order
extrapolation condition at the excision boundary is required for stable runs.

The paper is organized as follows: We describe the Kerr-Schild BH
spacetime and the scalar field equation in this background geometry in
Sec.~\ref{be}. Sec.~\ref{ivbc} is devoted to a discussion of initial
data, numerical algorithms, and different boundary conditions.  We
present our 3D numerical results, and compare them to very accurate 1D
results in Sec.~\ref{numr2}.  We summarize and discuss the
implications of our findings in Sec.~\ref{conc}.  We also include two
Appendices.  Appendix~\ref{staa} sketches the von Neumann 
stability analysis of the 1D centered and
upwind schemes.  Appendix~\ref{transn} describes the Lorentz
transformation of a scalar wave.  Throughout the paper we adopt
geometrized units with $G=c=1$.

%========================================
\section{Basic equations}
%========================================
\label{be}
%========================================
\subsection{Kerr-Schild form of the Schwarzschild spacetime}
%========================================

The ingoing Kerr-Schild form of the Kerr metric is given by 
\begin{equation} \label{ksf1}
ds^2=(\eta_{\mu\nu} + 2H\ell_\mu\ell_\nu)dx^\mu dx^\nu,
\end{equation}
(see \cite{matr,chas}), where $\mu$, $\nu$ run from $0$ to $3$,
$\eta_{\mu\nu}={\rm diag}(-1,1,1,1)$ is the Minkowski metric in
Cartesian coordinates, $H$ is a scalar function.  The vector
$\ell_\mu$ is null both with respect to $\eta_{\mu\nu}$ and $g_{\mu\nu}$,
\begin{equation}
   \eta^{\mu\nu}\ell_\mu\ell_\nu=g^{\mu\nu}\ell_\mu\ell_\nu=0,
\end{equation}
and we have $\ell^2_t=\ell^i\ell_i$. The spacelike hypersurfaces
extend smoothly through the horizon, and gradients near the horizon
are well-behaved.  Comparing the metric (\ref{ksf1}) with the ``ADM''
metric typically used in $3+1$ formulations, one identifies the lapse
function $\alpha$, shift vector $\beta_i$ and the spatial 3-metric
${}^{(3)}g_{ij}$ as
\begin{equation} \label{ksf2}
   \begin{array}{rcl}
              \alpha &=& 1/\sqrt{1+2H\ell^2_t}, \\
             \beta_i &=& 2H\ell_t\ell_i,\\
      {}^{(3)}g_{ij} &=& \eta_{ij} + 2H\ell_i\ell_j.
   \end{array}
\end{equation}
For the time-independent Schwarzschild spacetime (the 
``Eddington-Finkelstein form'' \cite{edda}) in Cartesian coordinates 
we have
\begin{equation}
   \begin{array}{rcl}
      H &=& M/r,\\
      \ell_\mu &=& (1,x_i/r),
   \end{array}
\end{equation}
where $M$ is the total mass-energy and $r^2=(x^1)^2+(x^2)^2+(x^3)^2$.
The Kerr-Schild metric~(\ref{ksf1}) is form-invariant under Lorentz
transformations.
Applying a constant Lorentz transformation $\Lambda$ (with boost
velocity ${\bf v}$ as specified in the background Minkowski spacetime)
to~(\ref{ksf1}) preserves the Kerr-Schild form,
but with transformed values for $H$ and $\ell_\mu$ (see \cite{matr})
\begin{equation}
   \begin{array}{rcl}
      x'^\alpha  &=& \Lambda^\alpha{}_\beta x^\beta,\\
      H(x^\alpha) &\rightarrow& H(\Lambda^{-1\alpha}{}_\beta x'^\beta),\\
      \ell_\mu' &=&\Lambda^\nu{}_\mu\ell_\nu(\Lambda^{-1\alpha}
      {}_\beta x'^\beta),\\
      g_{\mu\nu}' &=& \eta_{\mu\nu} + 2H\ell_\mu'\ell_\nu',
   \end{array}
\end{equation}

Now let $x^\alpha$ and $x'^\beta$ be the coordinates in the lab frame
$X$ and the comoving (with the BH) frame $X'$. For the
Eddington-Finkelstein system boosted in the $xy$-plane ($v_3=0$), in
the lab frame $X$ we have 
\begin{eqnarray}
     r^2 &=& \gamma^2(v_1\overline{x}+v_2\overline{y})^2 + \overline{x}^2
              +\overline{y}^2 +z^2,\nonumber\\
  \ell_t &=& \gamma[1-\gamma(v_1\overline{x}+v_2\overline{y})/r],\nonumber\\
  \ell_x &=& \overline{x}/r - v_1\ell_t, \label{ksf3}\\
  \ell_y &=& \overline{y}/r - v_2\ell_t,\nonumber\\
  \ell_z &=& z/r,\nonumber
\end{eqnarray}
where $\gamma=1/\sqrt{1-v^2}$ and $v^2=v_1^2+v_2^2$. $\overline{x}$
and $\overline{y}$ are defined as $\overline{x}\equiv x-v_1t$ and
$\overline{y}\equiv y-v_2t$. Under a boost the metric becomes
explicitly time dependent. Since the boost of the Schwarzschild
solution merely ``tilts the time axis'', we can consider all the
boosted $3+1$ properties at an instant $t=0$, in the frame which sees
the hole moving. Subsequent time $t$ simply offset the solution by an
amount $vt$. With Eqs.~(\ref{ksf3}) $\alpha$ and $\beta_i$ are defined
via Eq.~(\ref{ksf2}).

%========================================
\subsection{The scalar field equation}
%========================================

The field equation for a massless scalar field is 
\begin{equation} 
\label{sfe1}
   \Box\phi = 0,
\end{equation}
which can be expanded as 
\begin{equation}
   g^{\mu\nu}\phi_{;\mu;\nu}
   ={1\over\sqrt{-g}}(\sqrt{-g}g^{\mu\nu}\phi_{,\mu})_{,\nu}
   =0.
\end{equation}
We can decompose this second-order equation into two first-order-in-time 
equations by defining the auxiliary variable 
$\pi\equiv \Phi_{,t} - \beta^i\Phi_{,i}$, where $\Phi\equiv r\phi$, 
obtaining,
\begin{equation}
\label{3dfe}
   \left\{
   \begin{array}{rcl}
      \Phi_{,t} &=& \beta^i\Phi_{,i} + \pi,\\
      \pi_{,t}  &=& \beta^i\pi_{,i}  + {\cal F}.
   \end{array}
   \right.
\end{equation}
Here
\begin{eqnarray}
{\cal F} &=& - {1\over g^{00}}\left( {}^{(3)}g^{ij}\Phi_{,i,j} + A^i\Phi_{,i}
             + B\pi + {C\over r^2}\Phi \right),\nonumber\\
A^i &=&  g^{i\mu}{}_{,\mu} - {2\over r}g^{i\mu}(\ell_\mu + V_\mu)
       + g^{0\mu}\beta^i{}_{,\mu} + B\beta^i,\nonumber\\
B &=&  g^{0\mu}{}_{,\mu} - {2\over r}g^{0\mu}(\ell_\mu + V_\mu),\\
C &=&  g^{\mu\nu} \left( 3(\ell_\mu + V_\mu)(\ell_\nu + V_\nu) 
     - \eta_{\mu\nu} - V_\mu V_\nu \right) \nonumber\\
  && - rg^{\mu\nu}{}_{,\nu}(\ell_\mu + V_\mu), \nonumber\\
g^{0\mu}\beta^i{}_{,\mu} &=&  {\beta^i\over r} \left( \alpha^2 
                              (3\ell_t - 2\gamma) - 2\gamma - 3\beta^iV_i
                              + {1\over\ell_t}\right) \nonumber \\
                         && + {V^i\over r} (1-\alpha^2), \nonumber
\end{eqnarray}
and $V_\mu \equiv \gamma(-1,v_i)$ is the 4-vector of the velocity ${\bf v}$
in Cartesian coordinates.

In the comoving (non-boosted) frame, the scalar field
equation~(\ref{sfe1}) can be cast into a 1D radial equation which, for
spherical waves, reduces to
\begin{eqnarray}
   && \left( -(1+2H)\partial^2_{t'} + 4H\partial_{t'}\partial_{r'} +
   (1-2H)\partial^2_{r'} - {2\over {r'}}H\partial_{t'}
   \right. \nonumber \\ && \qquad\left. + {2\over r'}H\partial_{r'} -
   {2H\over {r'}^2} \right) \Phi=0.
\label{sfe2}
\end{eqnarray}
Eq.~(\ref{sfe2}) can again be decomposed into two first-order-in-time
equations
\begin{equation} \label{1dfe}
   \left\{
   \begin{array}{rcl}
      \Phi_{,t'}    &=& \beta^{r'}\Phi_{,r'} + \pi,\\
      \pi_{,t'}  &=& \beta^{r'}\pi_{,r'}  + f,
   \end{array}
   \right.
\end{equation}
where
\begin{eqnarray}
f &=& {1\over(1+2H)^2}\partial^2_{r'}\Phi + {2H\over r'(1+2H)^3}
	\partial_{r'}\Phi
      \nonumber \\
  && - {2H\over r'(1+2H)}\pi - {2H\over r'^2(1+2H)}\Phi,\\
\beta^{r'} &=& {2H\over 1+2H}. \nonumber
\end{eqnarray}
The characteristics (which can be derived from metric Eq.~(\ref{ksf1})) of 
the scalar field Eq.~(\ref{1dfe}) have speeds $-1$ and $(r'-2M)/(r'+2M)$. 

In the following sections we solve Eq.~(\ref{3dfe}) with a 3D code and
Eq.~(\ref{1dfe}) with a 1D code and compare the results. Since the 1D
code can be used with almost arbitrary resolution, we can effectively
compare our 3D results with an ``exact'' numerical solution.

%========================================
\section{Initial data, Numerical algorithms and boundary conditions}
%========================================
\label{ivbc}
%========================================

%========================================
\subsection{Initial value}
%========================================

As initial data for the scalar field in the comoving frame of the BH,
we choose a spherical 
Gaussian of width $\sigma$ centered at radius $r_0'$ 
\begin{equation}\label{phini}
\Phi(0,x',y',z')=\exp\left(-{(r_*'(r')-r_*'(r_0'))^2\over\sigma^2}\right)
\end{equation}
where $r_*' \equiv r' + 2M\ln(r'/2M-1)$ is the tortoise coordinate
(compare \cite{rezl}). 
For all calculations in this paper we choose $r_0'=10M$ and $\sigma=1M$.
According to (\ref{phini}) $\Phi$ vanishes on
the event horizon.  We also assume time symmetry at $t=0$
\begin{equation}\label{tsym}
\partial_t\Phi(0,x',y',z')=0.
\end{equation}
so that
\begin{equation}
\pi(0,x',y',z')=-\beta^{i'}\partial_{i'}\Phi(0,x',y',z').
\end{equation}

As the wave packet evolves in time, it splits, with one part of it
propagating outwards towards null infinity and the other propagating
inwards towards the horizon. Near the horizon, the wave undergoes
partial transmission and reflection.  We calculate the waveform of the
scattered scalar wave as observed at some fixed distance from the BH
and compare the 3D results with the ``exact'' 1D waveform.

%========================================
% Figure
%========================================
\begin{figure}[tb]
      \centering
      \leavevmode
      \epsfxsize=2.4in
      \epsfysize=1.8in
      \epsffile{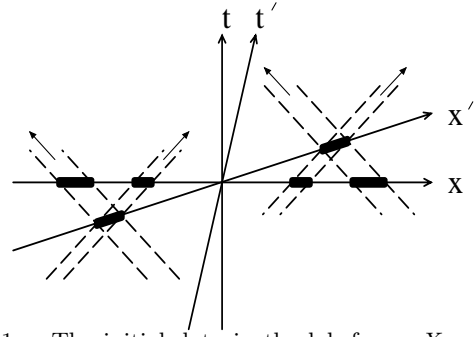}
\caption{
The initial data in the lab frame $X$ and in the comoving frame $X'$.
The bold solid line segments represent the wave packets (initial data) in 
different frames, 
the dashed lines indicate the propagations of the wave packets in spacetime.
The initial data in $X$ (at $t=0$) can be obtained by the future and past 
evolution of the scalar field with the initial data in $X'$ (at $t'=0$).
}
\label{fig1}
\end{figure}
%========================================

For the boosted cases, the initial data is different from the
non-boosted case due to the tilt of axes (see Fig.~\ref{fig1}).  The
frame $X'(t',{\bf x}')$ (comoving with the BH) moves along $x$-axis
with a boost velocity relative to the lab frame $X(t,{\bf x})$. The
initial data in $X$ can be derived by evolving the scalar field in
$X'$, followed by a Lorentz transformation. (Refer to
Appendix~\ref{transn} for details.)

%========================================
\subsection{Differencing scheme}
%========================================
\label{numr1}
%========================================

%========================================
% Table
%========================================
\begin{table*}[htb]
\begin{center}
\begin{tabular}{ccccccc}
recipe&advection term&advection term&inner&&stable?\\ \cline{5-7}
&outside BH&inside BH&boundary&$v=0$&$v=0.2$&$v=0.5$\\ \hline
AI&{\bf C}&{\bf C}&{\bf E}&Yes&Yes&No\\ \hline
AII&{\bf C}&{\bf U}&{\bf E}&Yes&Yes&Yes\\ \hline
BI&{\bf U}&{\bf U}&{\bf P}&Yes&No&No\\ \hline
BII&{\bf U}&{\bf U}&{\bf E}&Yes&Yes&Yes
\end{tabular}
\end{center}
\caption{Summary of recipes AI, AII, BI, and BII, 
and their stability properties for the three different boosts.
The boldface letters denote {\bf C}: centered scheme; 
{\bf U}: upwind scheme ; {\bf E}: third-order
extrapolation condition; {\bf P}: copying the time derivative of every
field at the boundary from the interpolated value just outside the 
excision zone along the
normal direction.  We analyzed the stability of these recipes for
static black holes ($v=0$) and boosts with speed $v=0.2$ and $v=0.5$.
Recipe BI, which is based on the implementation of {\protect
\cite{alcm2}}  passes the non-boosted case but fails in the boosted
cases.  Recipe AI passes the non-boosted case and the $v=0.2$ boosted
case but fails in the $v=0.5$ case.  Recipe AII and BII pass all
tests that we have performed.}
\label{reps}
\end{table*}
%========================================

%========================================
% Figure
%========================================
\begin{figure}[htb]
      \centering
      \leavevmode
      \epsfxsize=3.3in
      \epsffile{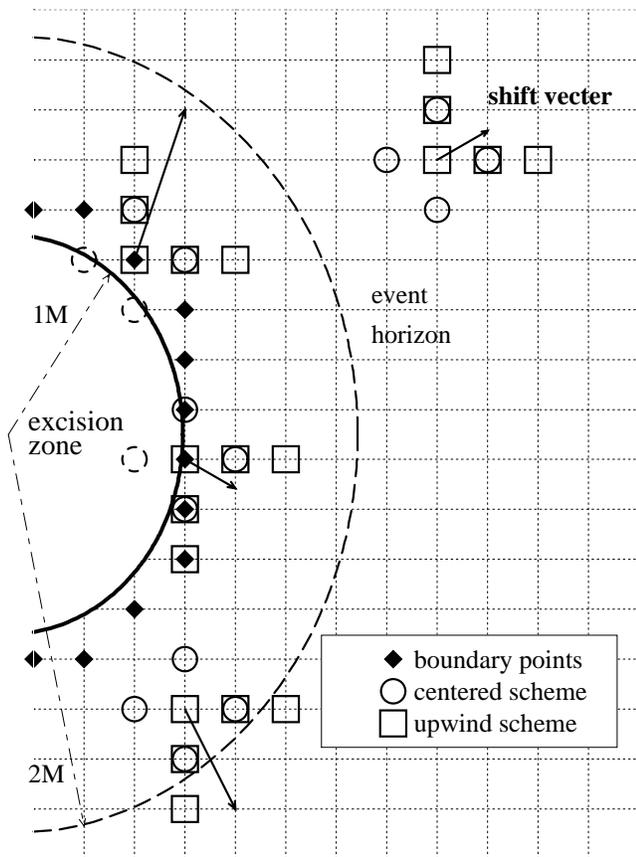}
\caption{
Schematic diagram of shift advection stencils in the computational
domain and around an excision zone.  The dashed circles inside the
excision zone are the grid points to which the extrapolated values are
assigned by using Eq.~(\ref{3rdex}).  }
\label{fig2}
\end{figure}
%========================================

We are looking for simple and robust recipes and focus, as in
\cite{alcm2}, on the shift terms in the field equation, which 
we regard as a generic ``advection'' terms (terms that look like
$\beta^i\partial_i$).  We apply both centered and upwind schemes to
these advection terms to test their stability.  All other
derivative terms are evaluated using centered differencing.

Consider the gridpoint with array indices $i$, $j$, and $k$ in the
$x$, $y$, and $z$ directions, respectively.  With centered
differencing, the first derivative of $\Phi$ with respect to $x$ at
this gridpoint is
\begin{equation}
   (\Phi_{,x})_{i,j,k}={1\over2\Delta x}[\Phi_{i+1,j,k}-\Phi_{i-1,j,k}].
\end{equation}
For the upwind scheme the second-order accurate first derivative along
$x$-direction is
\begin{eqnarray}
   (\Phi_{,x})_{i,j,k}&=&-{\nu_1\over2\Delta x}[\Phi_{i+2\nu_1,j,k}  
   \nonumber \\
   &&\qquad\qquad -4\Phi_{i+\nu_1,j,k} +3\Phi_{i,j,k}],
\end{eqnarray}
where $\nu_1$ is defined as
\begin{equation}
\nu_1\equiv{\beta^1\over|\beta^1|}=\left\{
\begin{array}{ll}
1\qquad&\mbox{for}\quad \beta^1>0,\\
-1\qquad&\mbox{for}\quad \beta^1<0.
\end{array}\right.
\end{equation}
We use analogous differencing in the $y$- and $z$-directions (see
Fig.~\ref{fig2}).

Different schemes can be used inside/outside the event horizon in each
recipe to further isolate where and how any instability arises.  In
this paper, we consider $4$ different recipes which we call AI, AII,
BI, and BII (see Table \ref{reps}).  Recipes AI, AII, and BII all use
the extrapolation inner boundary condition (see Sec.~\ref{innerbd}
below).  In recipe AI, the centered scheme is used for finite
differencing on the shift advection term everywhere. This scheme is
stable according to a van-Neumann stability analysis since $\beta^i$
is never greater than unity in the Kerr-Schild metric (refer to
Appendix~\ref{staa} for a stability analysis).  In AII, the upwind
scheme is used inside the BH and the centered scheme is used outside
the BH for the shift term.  In BII, upwind differencing is used
everywhere. 

Recipe BI, which is based on the implementation of \cite{alcm2},
differs from the other schemes in its excision boundary condition.
Alcubierre and Br\"ugmann \cite{alcm2} use a cubic excision zone and
copy the time derivative of every field at a gridpoint just inside the
excision zone from the neighboring gridpoint just outside.  Our
excision zone is spherical with radius $r' = M$ (in the comoving
frame).  To generalize the scheme of \cite{alcm2} we interpolate
neighboring gridpoints to the normal on the surface of the excised
region.  More specifically, for a gridpoint $(i,j,k)$ just inside the
excised region we take its nearest neighbors along the coordinate axes
away from the center of the black hole, say $(i+1,j,k)$, $(i,j+1,k)$
and $(i,j,k+1)$.  These three points define a plane, and we
interpolate to the intersection of this plane with the normal on the
surface of the excised region.  If one of these three neighbors is
still in the excised region, we project the normal into the plane
spanned by the remaining two coordinate axes, and do the interpolation
there; if two of the neighbors are inside the excised region we
directly copy the remaining third point.  If all three points are
inside the excised region we repeat the procedure with the three
points $(i+1,j+1,k)$, $(i+1,j,k+1)$ and $(i,j+1,k+1)$; if that is also
not successful we finally copy the point $(i+1,j+1,k+1)$.  
This particular algorithm is only one of many possibilities.  We have
experimented with a number of other schemes, and have found that the
stability properties do not depend on the details of this
implementation. 
Like BII, BI also uses the upwind scheme to compute the shift advection terms
everywhere.

%========================================
\subsection{Outer boundary conditions}
%========================================

At the outer boundary, we impose an outgoing wave boundary condition.
In this approximation, we assume that the functions are of the form
$\Phi=f(\lambda t-R)$ (since $\Phi=r\phi$), where
$\lambda=(R-2M)/(R+2M)$ is the outgoing characteristic speed. The
values $f(t+\Delta t,R)$ of the grid points at the outer boundary are
updated by using the value $f(t,R-\lambda\Delta t)$ with second-order
interpolation. In the boosted cases, the characteristics $\lambda$ at
the outer boundary are derived from the relativistic addition of the
characteristic in the comoving frame and the boost velocity.

This boundary condition provides a stable outer boundary provided that
the outer boundary is placed at a sufficiently large distance. With
the outer boundary at finite radii, as in our cases, some reflected
waves are created at the boundary. (We find that reflection waves with
larger amplitude will appear if the true characteristic speed
$\lambda$ in $f$ is replaced by $1$) The amplitude of the reflected
wave decreases as resolution is increased.  In boosted cases, the
effect of the inaccurate outer boundary condition cannot be ignored as
the BH approaches the outer boundary.  For our tests, however, the
reflected waves from the boundary are small perturbations of the field
and do not affect the stability of different recipes, only the
accuracy.

%========================================
\subsection{Inner boundary conditions}
%========================================
\label{innerbd}
%========================================

%========================================
% Figure
%========================================
%\begin{figure*}[htb]
%\begin{center}
%   \leavevmode
%   \epsfxsize=3in
%      \epsffile{fig3a.ps}
%   \epsfxsize=2in
%      \epsffile{fig3b.ps}
%\end{center}
%\caption{
%Convergence tests for the radial 1D codes (stationary BH).
%All the curves show the leading (second) order truncation error defined as
%${\rm Err}_{\Delta r} = \Phi_{\Delta r/2} -\Phi_{\Delta r}$ for
%the amplitude of a spherical wave $\Phi$ at the check point $r=14M$.
%The left panel is the result of the 1D code using recipe AI.
%The right panel is the result of the 1D codes using recipe BII.
%}
%\label{fig3}
%\end{figure*}
%========================================

At the inner boundary we use a singularity excising method (see,
e.g.~\cite{seie,annp,schm2,schm,bonc,abra,marr}). In general dynamic
spacetimes, the location of the apparent horizon is not known {\it a
priori}, and must be computed at each time step with an ``apparent
horizon finder'' (e.g.~\cite{baut3,thoj,gomr2}). This is not the case
for the present static Schwarzschild background since the location of
the apparent (and event) horizon is known at all times \cite{fora}.
On the other hand, it is not necessary to know the exact location of
the apparent horizon to implement the singularity-excising method,
provided the excision zone is sufficiently small that one can be
confident that it lies entirely inside the horizon.  We choose to
excise the region with radius $r'=1M$ from the center of BH where $r'$
is the radial distance measured in the comoving frame $X'$.  The
simplest inner boundary condition obtains, by extrapolation, the field
variables at gridpoints just inside the masked region from gridpoints
just outside this region.  These values can then be used in a centered
evolution scheme to update the field variables at gridpoints just
outside the excision zone. We have implemented such a boundary
condition using a third-order extrapolation scheme
\begin{equation}\label{3rdex}
   f_{j-\nu} = 4f_j - 6f_{j+\nu} + 4f_{j+2\nu} - f_{j+3\nu},
\end{equation}
in all recipes except BI.  Here $j$ is a gridpoint just outside the
excised region, and, with $\nu$ either $+1$ or $-1$, $j - \nu$ is just
inside (see Fig.~\ref{fig2}).  We apply this algorithm along whichever
axis the second derivative is taken, for example along the $x$-axis
for a second derivative with respect to $x$.  With this extrapolation,
the second spatial derivatives using centered differencing are
second-order accurate.  This is equivalent to using a second-order
one-sided differencing scheme.  These boundary conditions do not
violate causality since no information is extracted from within the
excision zone.  This prescription is simple to implement and does not
require special assumptions on the behavior of the variables in the
proximity of the excision zone. A similar implementation has proven to
be stable for wave propagation in $2+1$ dimensions on a flat spacetime
\cite{abra}.

In recipe BI, we generalize the implementation of \cite{alcm2} and
copy the time derivative of every field variable at the boundary from
interpolated values just outside the excised region as described in
Sec.~\ref{numr1}.  As we will show in Sec.~\ref{numr2}, this inner
boundary condition will produce stable results in the non-boosted BH
case, but results in an instability when the BH moves and grid points
emerging from the excision zone must be assigned by extrapolation.

%========================================
\subsection{Extrapolation to newly emerging grid points}
%========================================

%========================================
% Figure
%========================================
\begin{figure*}[htb]
   \centering
   \leavevmode
   \begin{tabular}{cc}
   \epsfxsize=3.3in
   \epsfysize=2.2in
      \epsffile{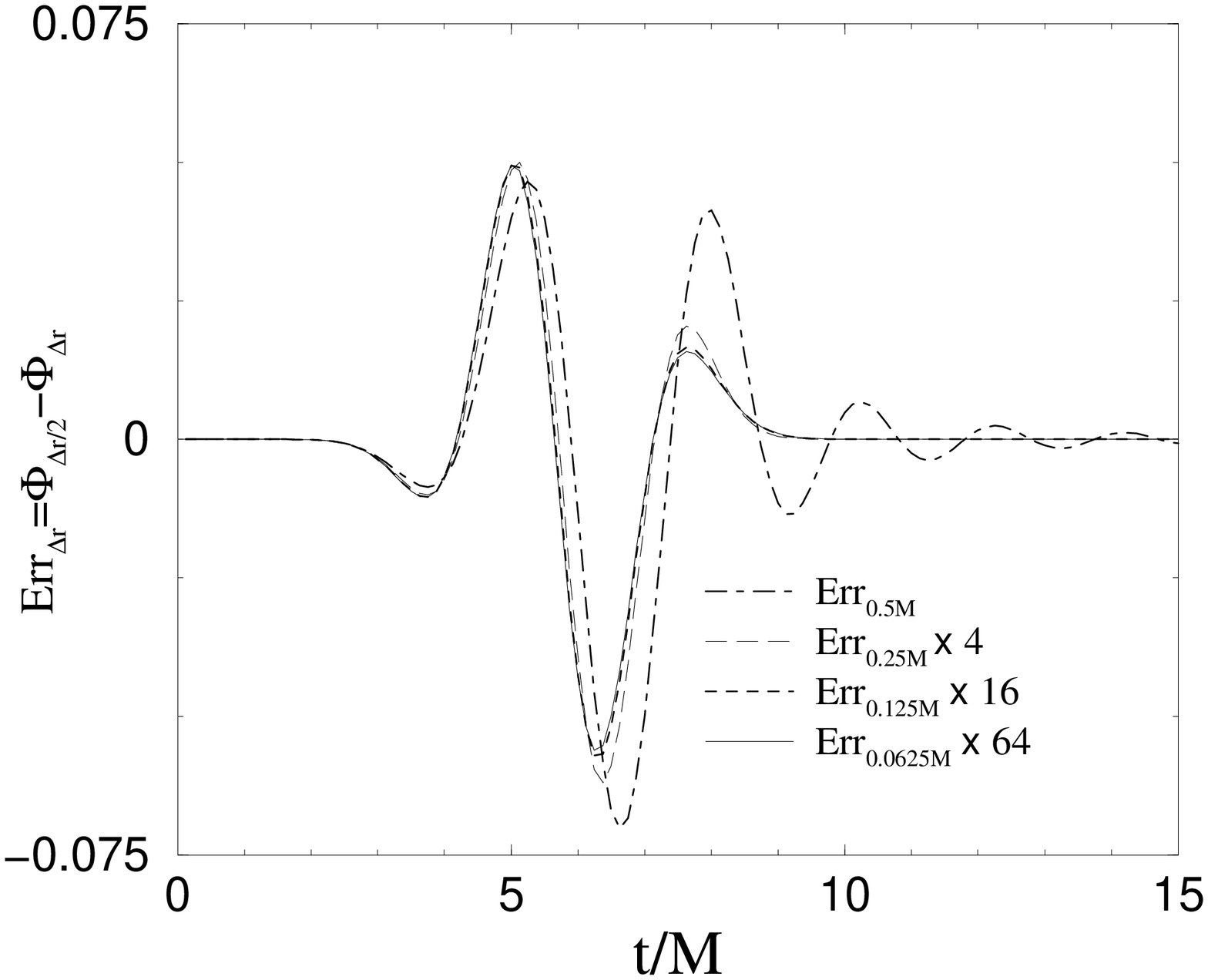}&
   \epsfxsize=3.3in
   \epsfysize=2.2in
\epsffile{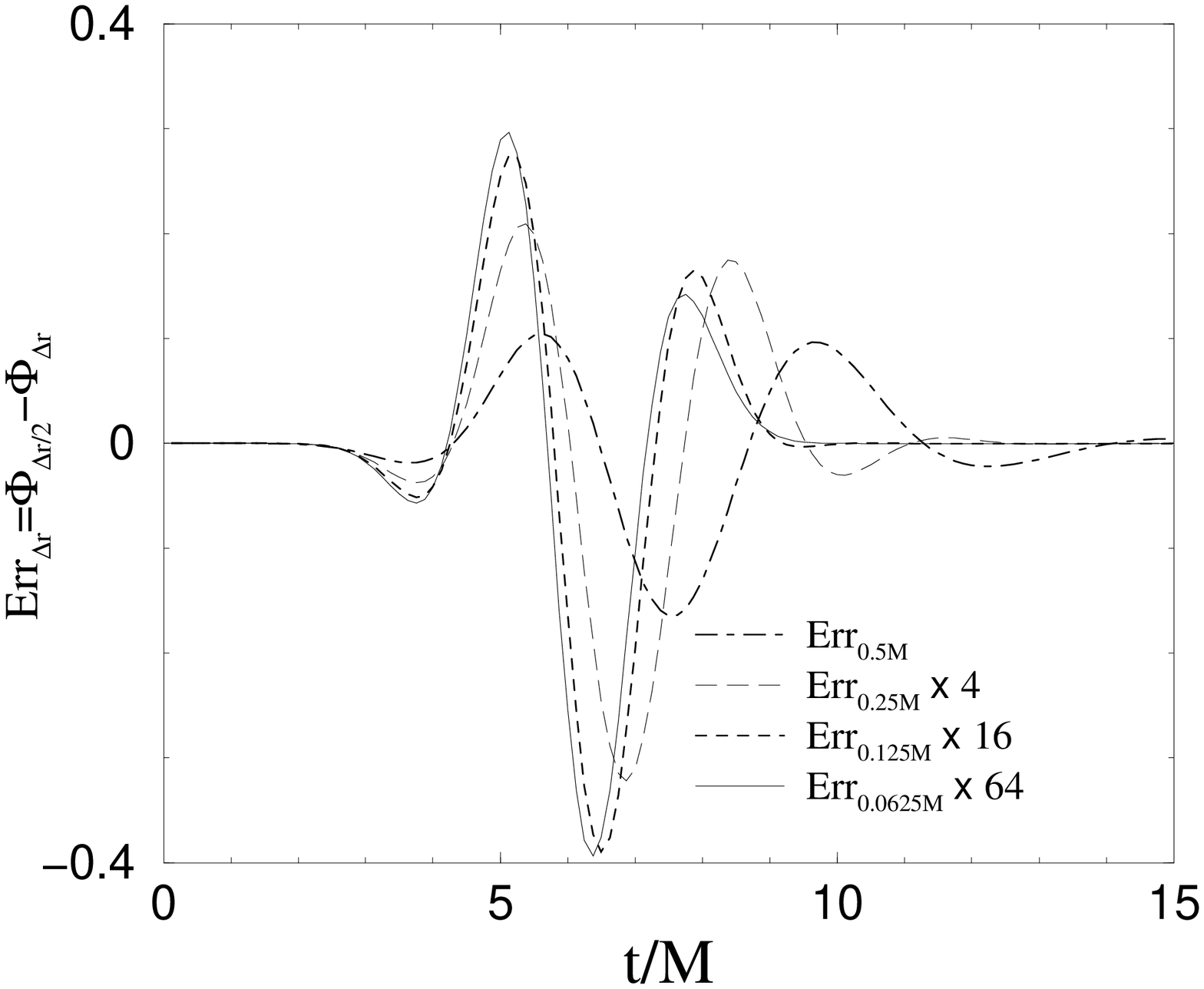}
   \end{tabular}
\caption{Convergence tests for our 1D evolution calculations, using
recipe AI (left panel) and BII (right panel).  To demonstrate second
order convergence, we plot the appropriately rescaled finite
difference error ${\rm Err}_{\Delta r} = \Phi_{\Delta r/2}
-\Phi_{\Delta r}$ at the check point $r=14M$ for various resolutions
$\Delta r$.  }
\label{fig3}
\end{figure*}
%========================================

We use a second-order extrapolation scheme to set values at gridpoints
that are newly emerging from the excision zone as the BH travels
through the numerical grid.  If available, we carry out the
extrapolation along the $z$-axis, because the BH is boosted in the
$xy$-plane, and the extrapolated values along the $z$-direction are
least affected by numerical errors from previous extrapolations in the
evolution.  If the points needed for this extrapolation are themselves
excised, we instead average between extrapolations along the $x$ and
$y$ directions or use only one of the two if the other one is not
available (because the necessary points are excised).  If neither one
is available we continue through the following hierarchy of preferred
extrapolation directions: $z$, then $x$ and $y$, then $xz$ and $yz$,
finally the $xy$- and $xyz$-directions.  We have experimented with
other extrapolation prescriptions and have found that the stability
properties of the code are fairly insensitive to the details of this
scheme.

%The value assigned at a grid point is a
%weighted average of extrapolations from the $7$ different directions
%$z$, $x$, $y$, $xz$, $yz$, $xy$, and $xyz$ to the point in
%question. For the extrapolation we weigh grid points in the
%$z$-direction most heavily, then $x$- and $y$-, then $xz$- and $yz$-,
%finally the $xy$- and $xyz$-directions.  The weight given to the
%extrapolation in the $z$-direction is heaviest since the BH is boosted
%in $xy$-plane, and the extrapolated values along the $z$-direction are
%least affected by numerical errors from previous extrapolations in the
%evolution.  If an extrapolation direction is not available (e.g., the
%points in that direction are themselves inside the excision zone),
%then it is bypassed in the algorithm.  We have experimented other
%extrapolation prescriptions, many of which are also stable, but we
%find empirically the one defined above gives the most accurate
%results.

%========================================
\section{numerical results}
%========================================
\label{numr2}
\begin{figure*}[htb]
   \centering
   \leavevmode
   \epsfxsize=6.6in
   \epsffile{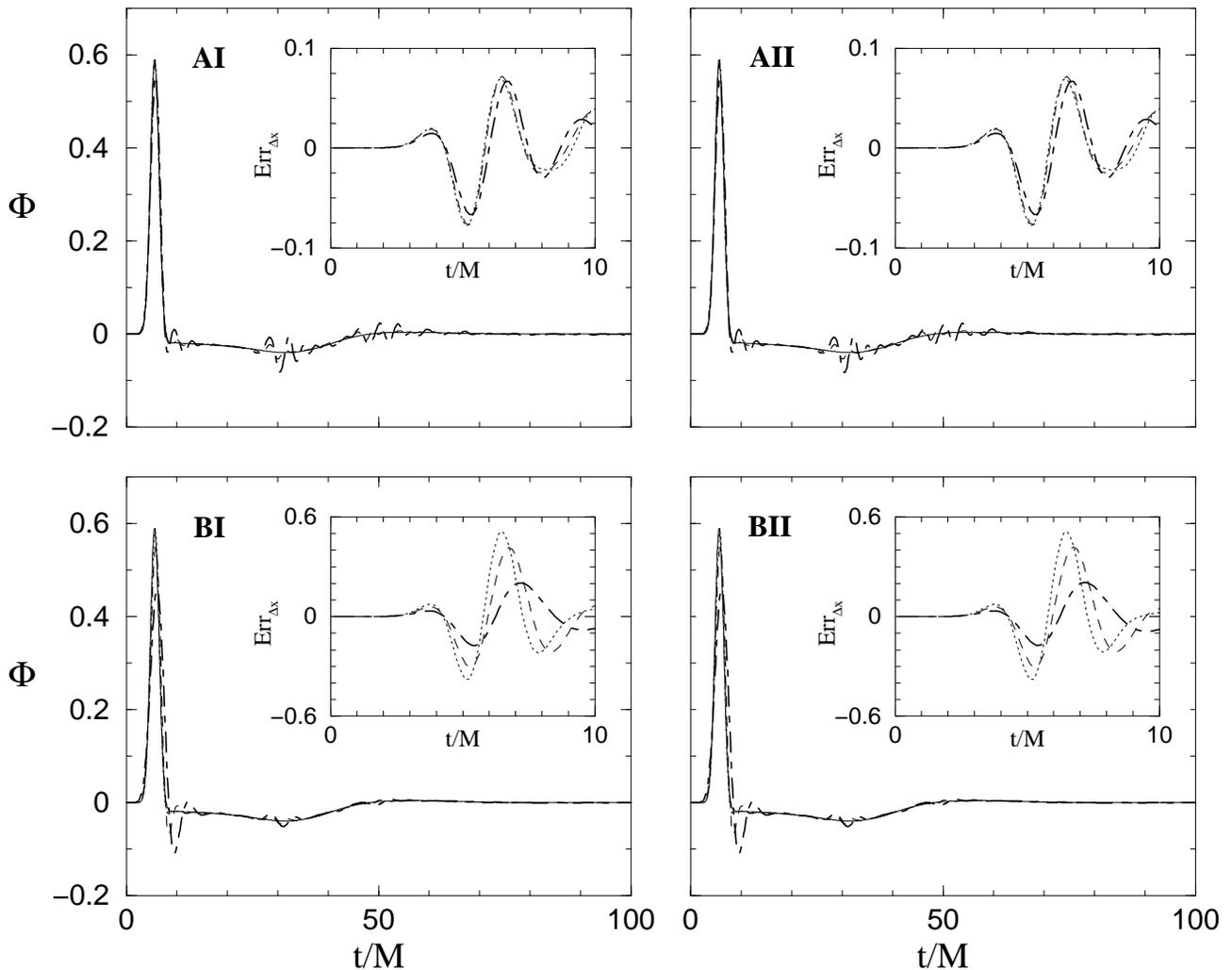} 

\caption{ The scalar field $\Phi$ as a function of $t$ at radius
$r=14M$ in the non-boosted 3D case.  In each panel, the solid curve is
the 1D solution, while the dotted, dashed and dot-dashed line are 3D
solutions computed for grid widths $\Delta =0.125M$, $\Delta =0.25M$,
and $\Delta =0.5M$, respectively.  We also include convergence tests
for each recipe in the insets, where the dotted line is $16\times{\rm
Err}_{0.125M}$, the dashed line is $4\times{\rm Err}_{0.25M}$, and the
dot-dashed line is ${\rm Err}_{0.25M}$.  Here the finite difference
error is computed from ${\rm Err}_{\Delta x}\equiv\Phi_{\Delta
x}-\Phi_{\rm 1d}$.  }
\label{fig4}
\end{figure*}
%========================================

We numerically solve both the 1D and 3D scalar wave
equations~(\ref{3dfe}) Eq.~(\ref{1dfe}) using an iterative
Crank-Nicholson scheme with two corrector steps~\cite{teus}.  Since
the 1D code can be run with an essentially arbitrary number of radial
grid points and hence essentially arbitrary accuracy, we use this
solution as the ``exact'' solution for comparisons.

We compare the results from the 3D simulations using the four recipes
(see Table.~\ref{reps}) for both the non-boosted and boosted
cases. There are two sub cases of the boosted case: a ``slow'' boost
speed $v=0.2$ and a ``fast'' boost speed $v=0.5$.  The results for
both boosted cases are summarized in Table.~\ref{reps}, but we will
discuss in detail only the $v=0.5$ boost case.

In each case, we compare the waveforms with the 1D result for three
uniform grid resolutions $\Delta=0.5M$, $\Delta =0.25M$,and $\Delta
=0.125M$. We choose a time step $\Delta t = \Delta / 4$ in all cases,
and assume equatorial symmetry across the $z=0$ plane.

In the non-boosted case, the computational domain is $32M\times
32M\times 16M$ ($-16M<x<16M$,$-16M<y<16M$,$0<z<16M$).  In the $v=0.5$
case, the domain is extended to $48M\times 48M\times 24M$
($-20M<x<28M$,$-24M<y<24M$,$0<z<24M$).  A larger spatial domain is
needed in the high speed case to observe the BH's motion before it
moves out of the computational domain.  In each 3D case the code with
$\Delta=0.125M$ resolution is run only to $t=10M$ to check its
stability and convergence; the lower resolution non-boosted cases are
run until $t=100M$, and the lower resolution boosted cases are run
until $t=48M$.

%========================================
\subsection{1D result}
%========================================

We verify the second-order accuracy of our 1D codes in
Fig.~\ref{fig3}. We compare the amplitude difference between the
results observed from the check point $r=14M$ by continual doubling of
the grid resolution.  We present results for recipe AI in the left
panel of Fig.~\ref{fig3} and for BII in the right panel.  Comparing
the scales of ${\rm Err}_{\Delta r}$ in the two panels shows that,
while both scheme are second order accurate, the centered scheme of
recipe AI converges much faster than the more diffusive upwind scheme
of recipe BII.  We find very similar results in the 3D codes below.

%========================================
\subsection{3D results in co-moving coordinates}
%========================================
 
%========================================
% Figure
%========================================
%\begin{figure*}[htb!]
%\centering
%\begin{center}
%\leavevmode
%\epsfxsize=3in
%\epsffile{fig5a.ps}
%\leavevmode
%\epsfxsize=3in
%\epsffile{fig5b.ps}
%\end{center} 
%\caption{
%The scalar field $\Phi$ as a function of time for a boost with $v=0.5$
%for checkpoints at ($14M$,$0$,$0$), ($0$,$14M$,$0$), and
%($-14M$,$0$,$0$) (from left to right).  The inset in each panel is the
%blowup of the waveform for early times.  The first and second rows are
%the results for recipes AII and BII respectively.  In each panel, the
%solid curve is the 1D solution, the dotted line, the dashed line, and
%the dot-dashed line are 3D solutions for grid resolutions
%$\Delta=0.125M$, $\Delta=0.25M$, and $\Delta=0.5M$.  In the third
%column, for checkpoint ($-14M$,$0$,$0$), we only show the two coarser
%resolutions.}
%\label{fig5}
%\end{figure*}
%========================================

%========================================
% Figure
%========================================
\begin{figure*}[htb!]
   \centering
   \begin{tabular}{c}
   \epsfxsize=6.6in
      \epsffile{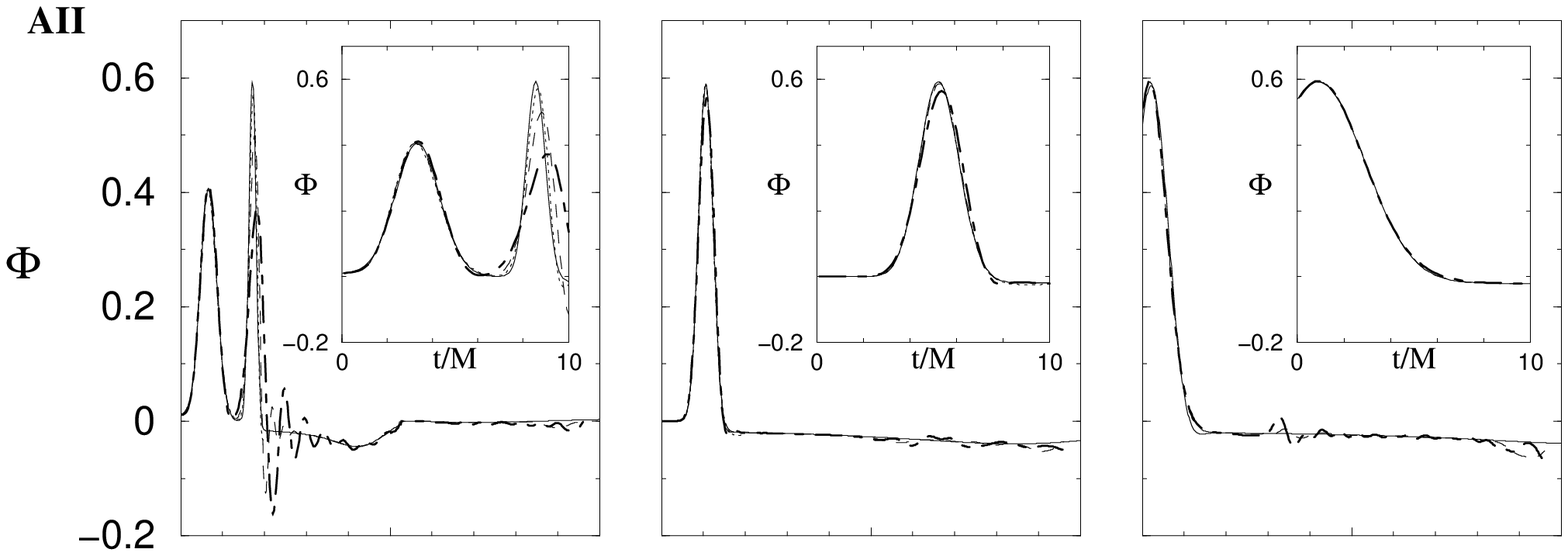}\\
   \epsfxsize=6.6in
      \epsffile{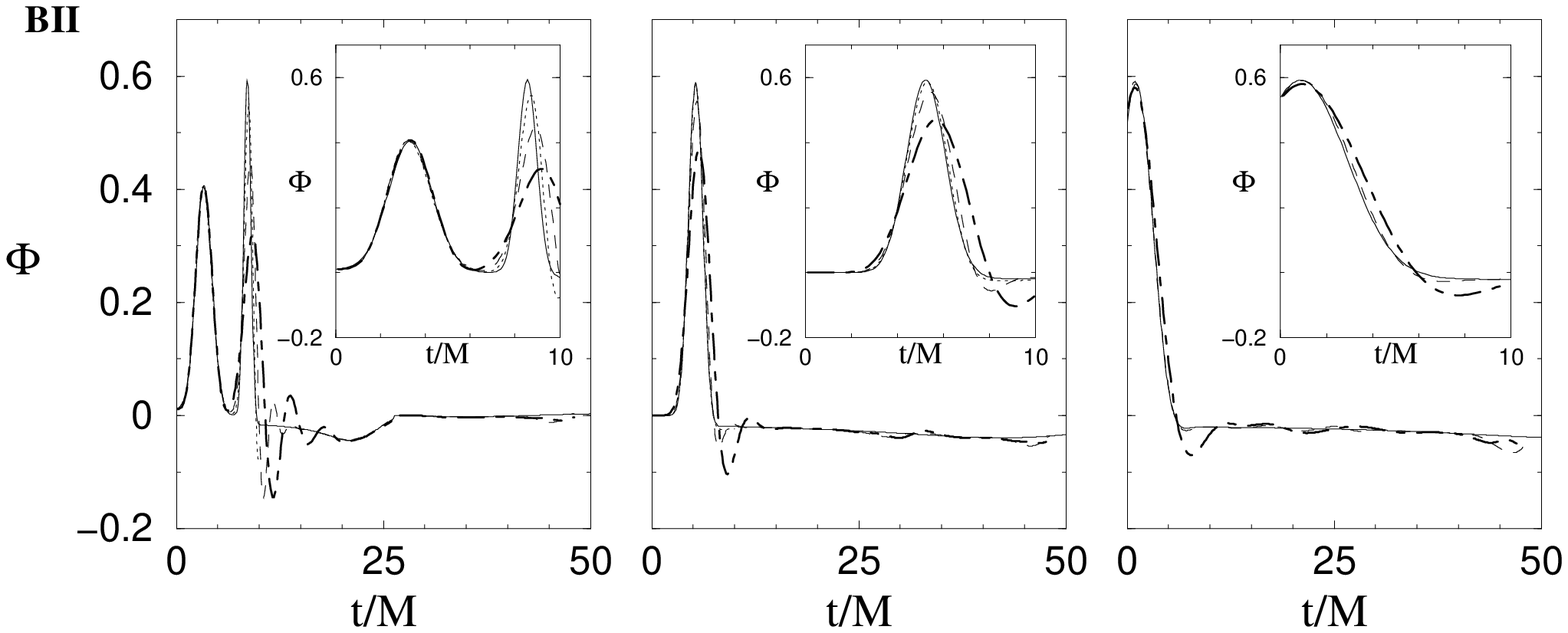}
   \end{tabular}
\caption{
The scalar field $\Phi$ as a function of time for a boost with $v=0.5$
for checkpoints at ($14M$,$0$,$0$), ($0$,$14M$,$0$), and
($-14M$,$0$,$0$) (from left to right).  The inset in each panel is the
blowup of the waveform for early times.  The first and second rows are
the results for recipes AII and BII respectively.  In each panel, the
solid curve is the 1D solution, while the dotted, dashed and
dot-dashed line are 3D solutions for grid resolutions
$\Delta=0.125M$, $\Delta=0.25M$, and $\Delta=0.5M$.  In the third
column, for checkpoint ($-14M$,$0$,$0$), we only show the two coarser
resolutions.}
\label{fig5}
\end{figure*}
%========================================

%========================================
% Figure
%========================================
\begin{figure*}
   \centering
   \leavevmode
   \epsfxsize=2.7in
   \epsfysize=1.8in
   \begin{tabular}{cc}
   \epsfxsize=2.7in
   \epsfysize=1.8in
      \epsffile{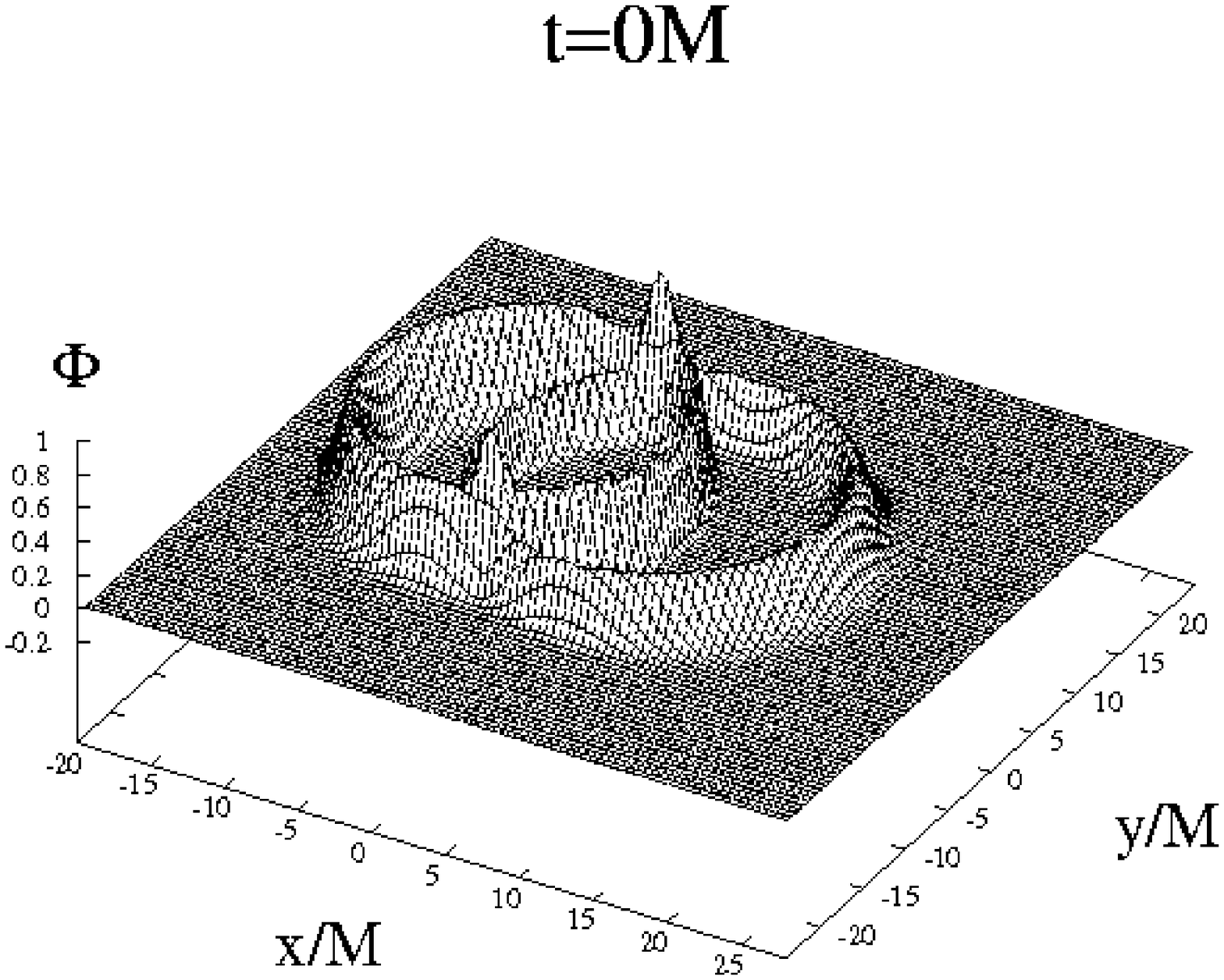}&
   \epsfxsize=2.7in
   \epsfysize=1.8in
\epsffile{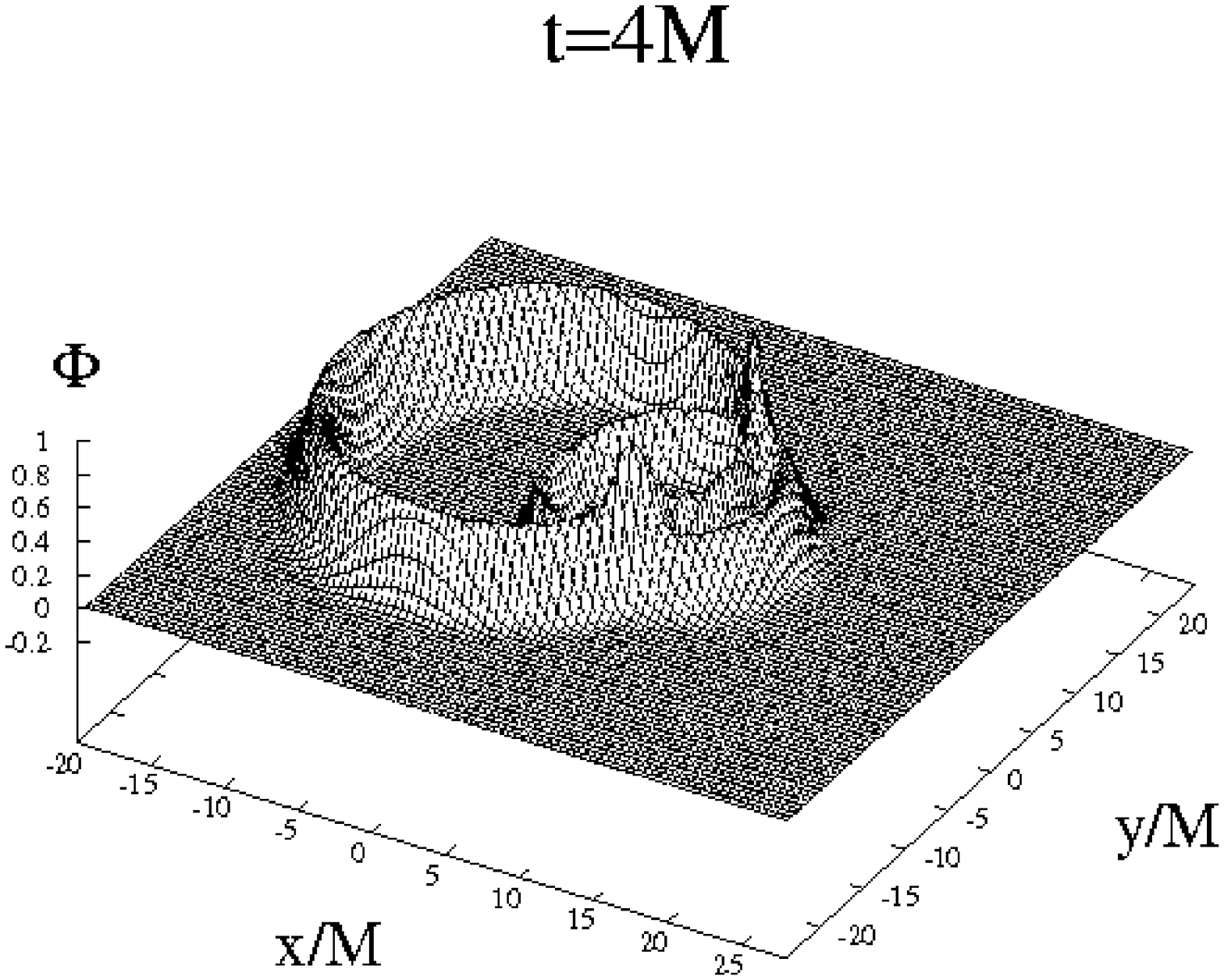}\\
   \epsfxsize=2.7in
   \epsfysize=1.8in
      \epsffile{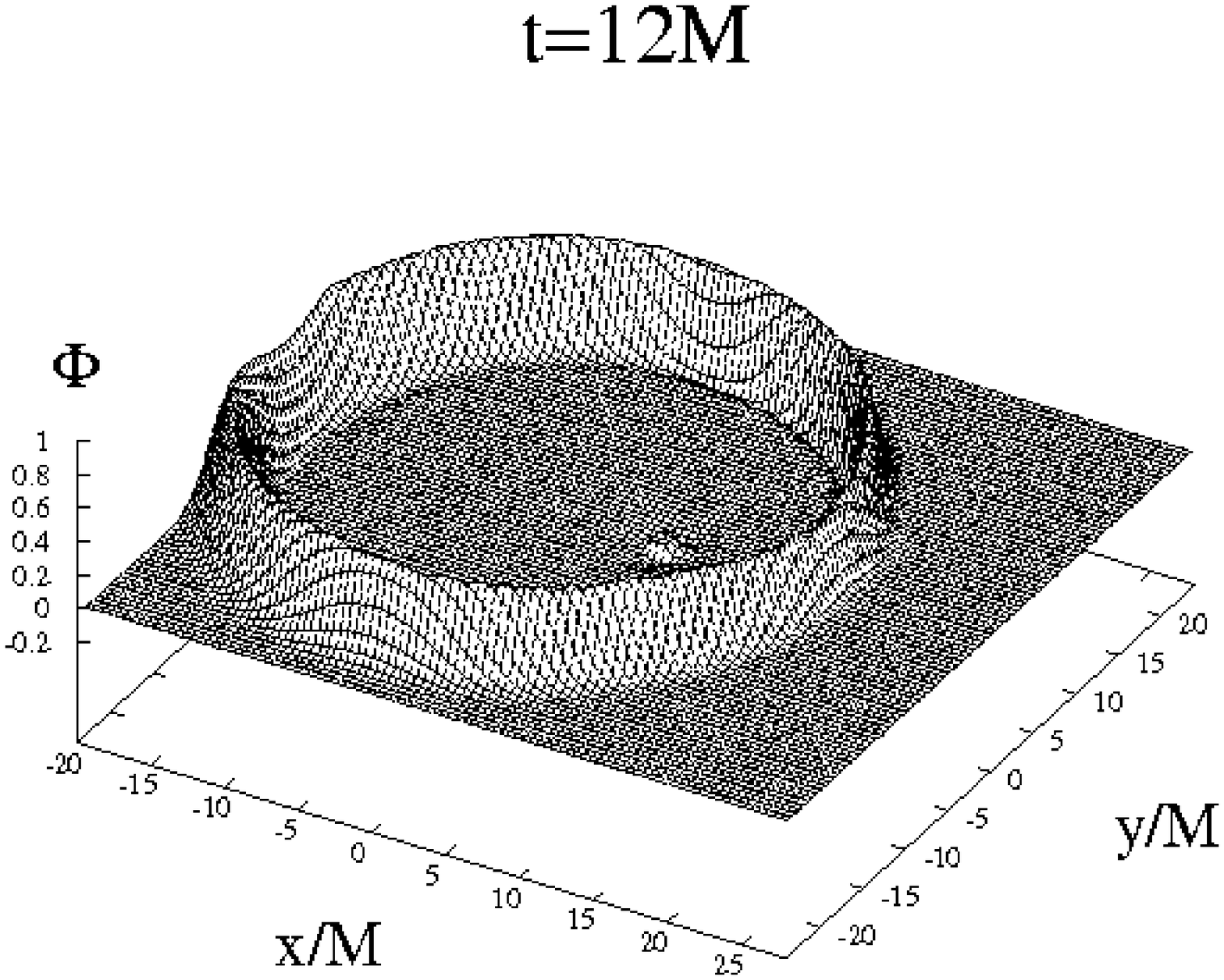}&
   \epsfxsize=2.7in
   \epsfysize=1.8in
\epsffile{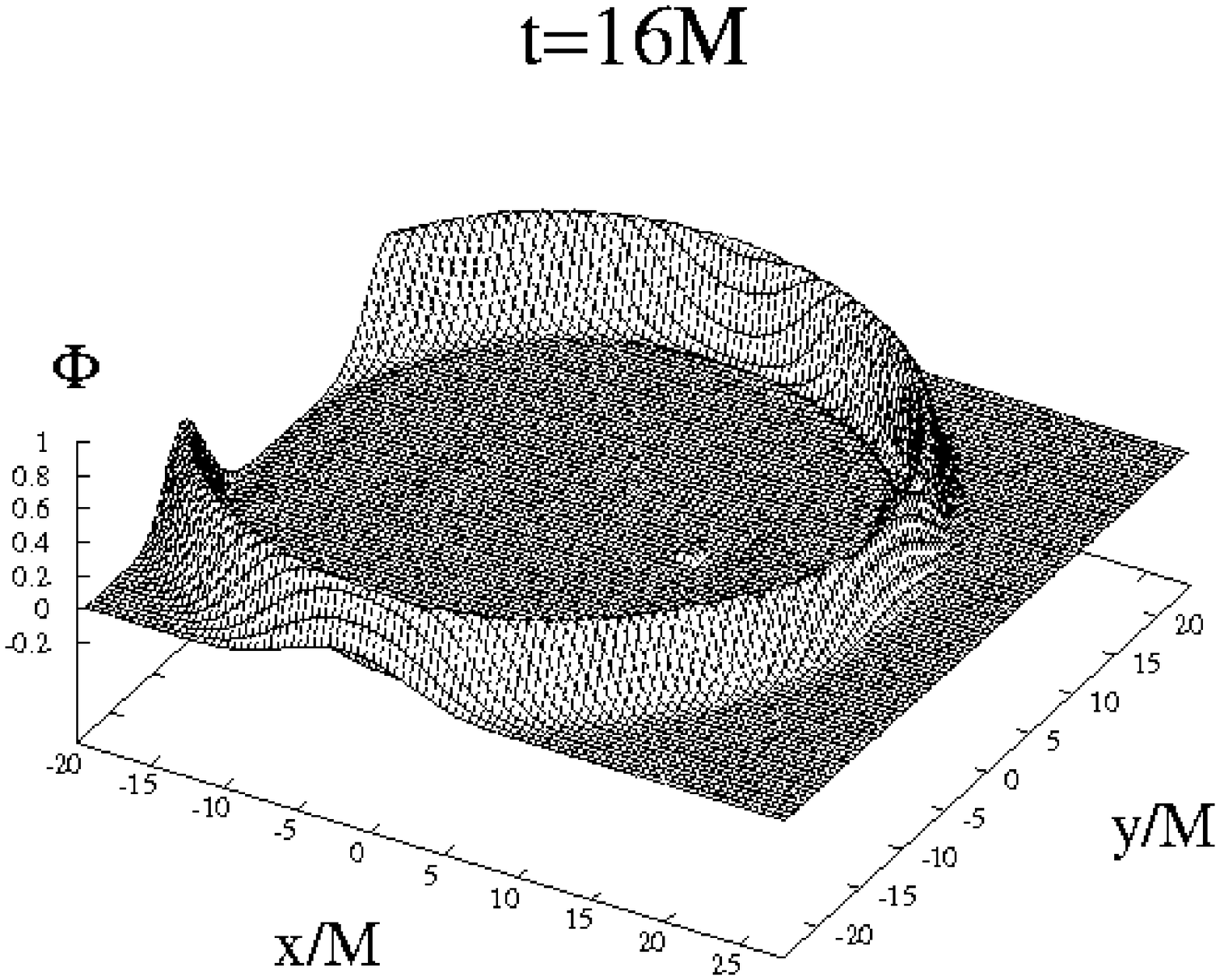}\\
   \epsfxsize=2.7in
   \epsfysize=1.8in
      \epsffile{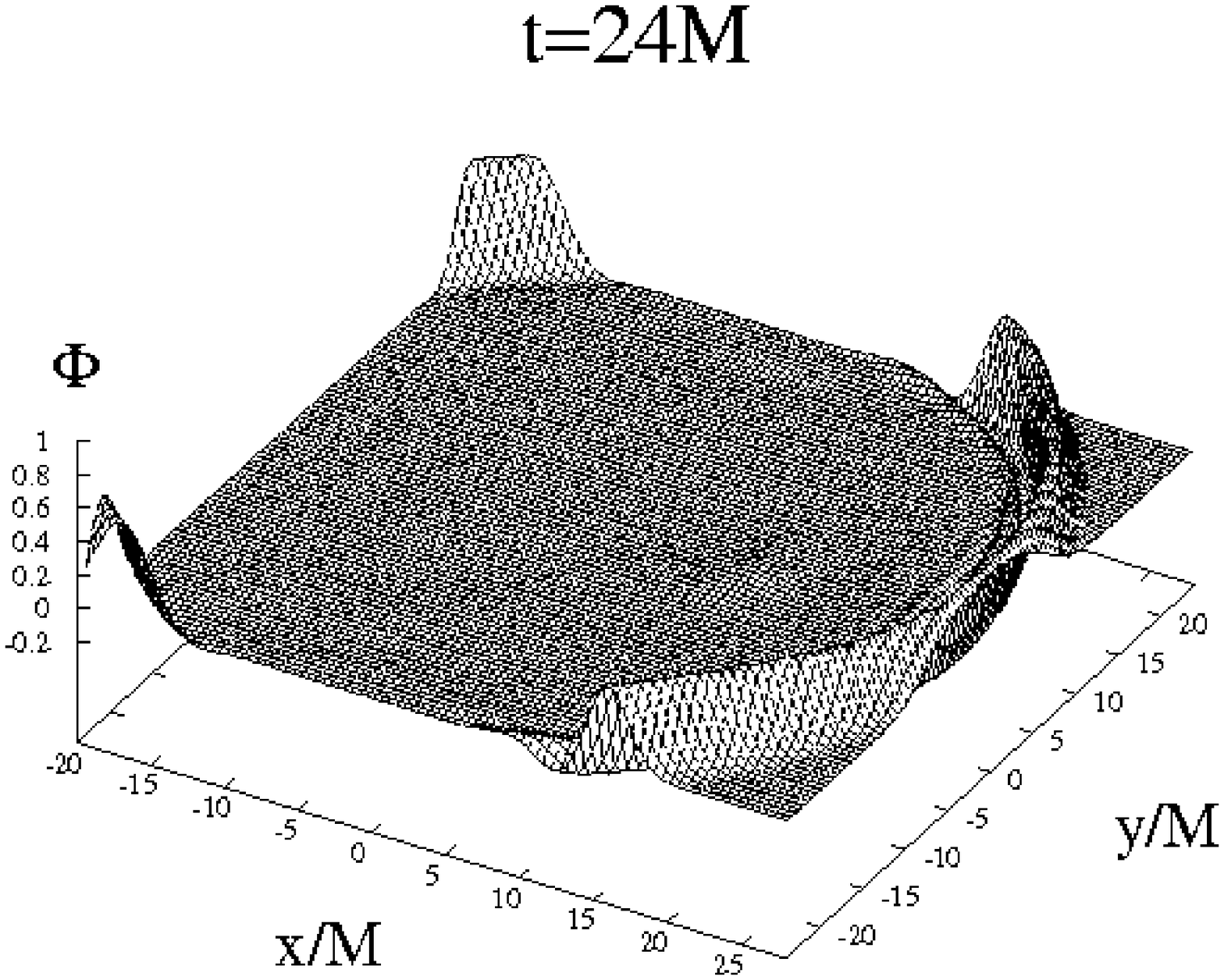}&
   \epsfxsize=2.7in
   \epsfysize=1.8in
\epsffile{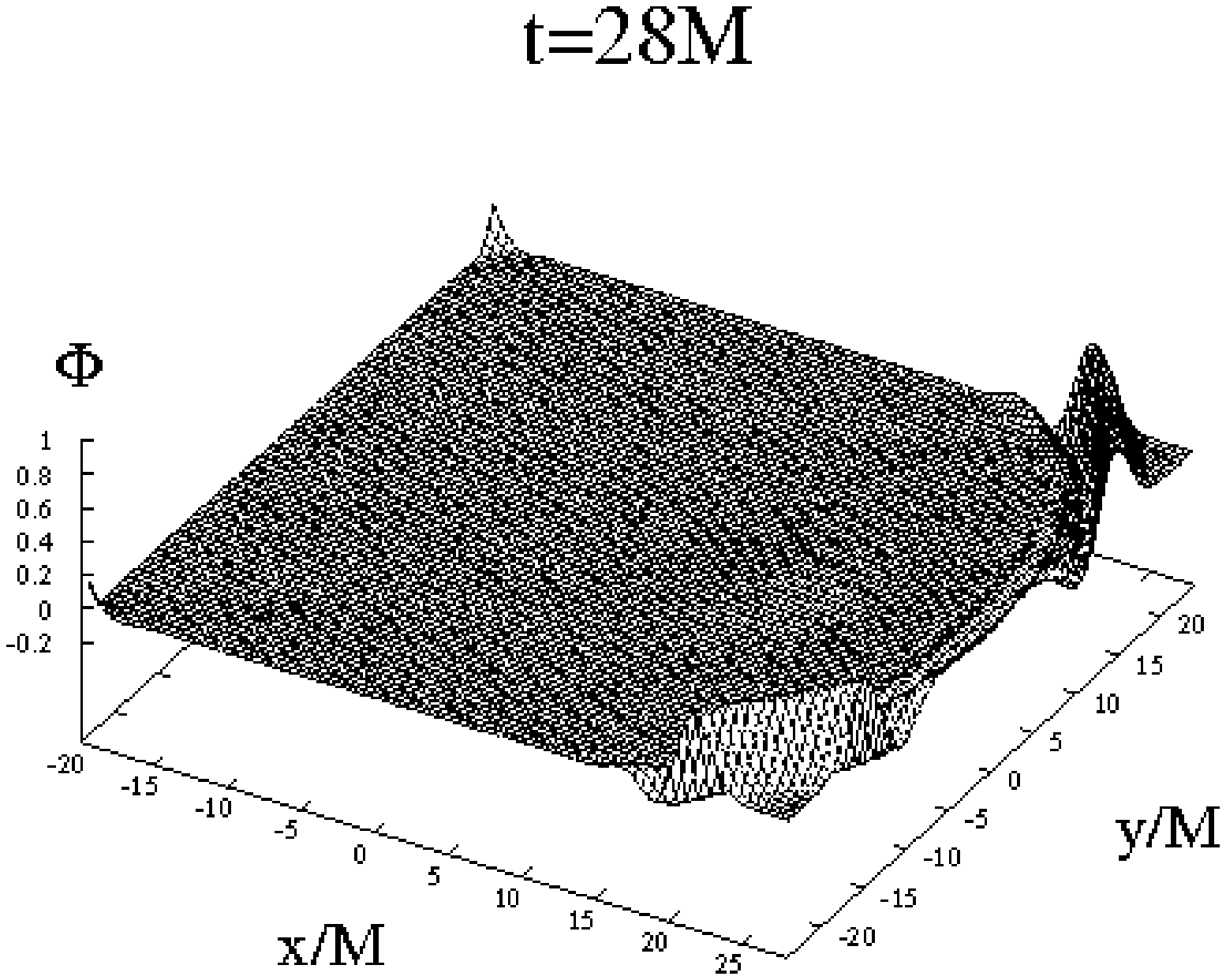}\\
   \epsfxsize=2.7in
   \epsfysize=1.8in
      \epsffile{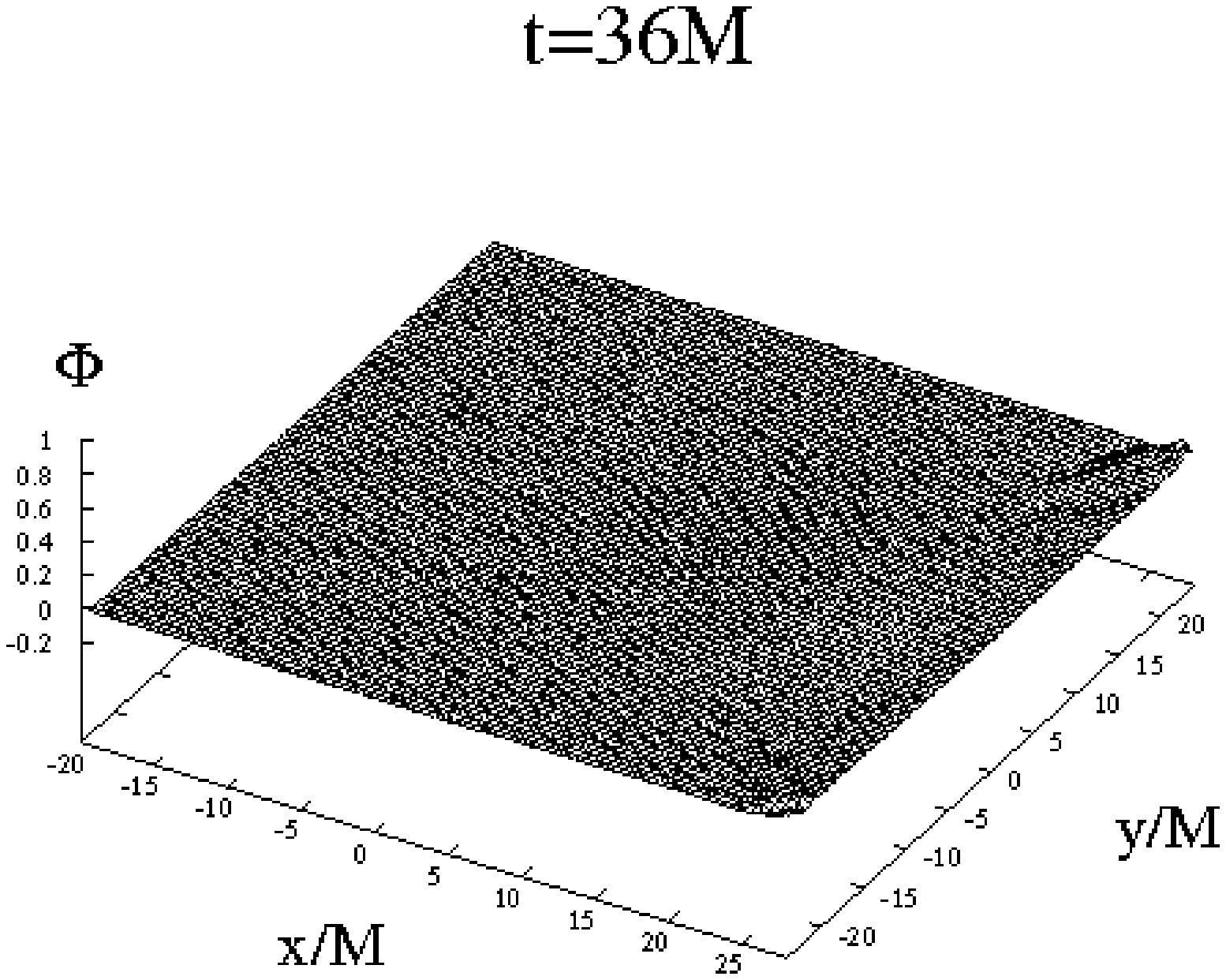}&
   \epsfxsize=2.7in
   \epsfysize=1.8in
\epsffile{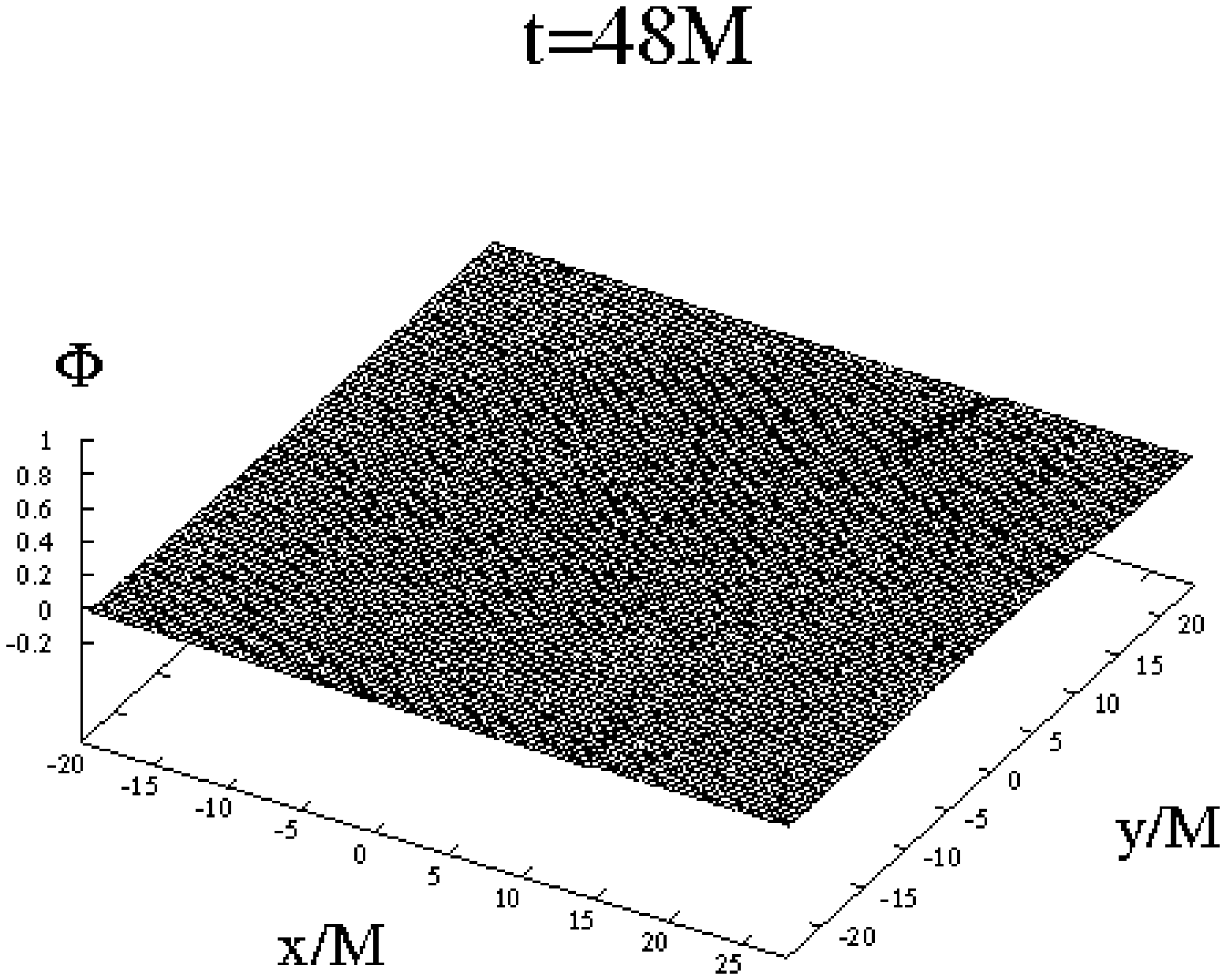}
   \end{tabular}
\caption{ The snapshots of the scalar field as evolved by recipe BII
in the boosted $v=0.5$ case ($z=0.125M$).}
\label{fig6}
\end{figure*}
%========================================

We first study the four recipes AI, AII, BI and BII for the
non-boosted (comoving) case. All four recipes give stable results and
converge to the 1D result, as shown in Fig.~\ref{fig4}. The stability
of recipe BI is consistent with the conclusion of \cite{alcm2} in which
the recipe gives a stable and accurate solution in a non-boosted
distorted BH.

In Fig.~\ref{fig4} the first peak and the first minimum in the
waveform curves are produced by the part of the wave packet which
moves outwards and leaves the computational grid. The minimum observed
at $t\approx10M$ is a numerical artifact and converges away for
increasing grid resolution.  We demonstrate the second order
convergence of the algorithms in the insets in Fig.~\ref{fig4}.  For
the upwind schemes of recipes BI and BII the convergence is much
slower than for the centered schemes of recipes AI and AII, as
expected from the 1D results (see Fig.~\ref{fig3}).  There are several
bumps at $t\approx10M$, $t\approx30M$, and $t\approx50M$.  These are
the reflection waves coming from the outer boundaries due to the
approximate outer boundary condition.  We find that these bumps can
always be diminished significantly by extending the outer boundary to
a larger radius and/or increasing the resolution.

The waveforms in Fig.~\ref{fig4} also show that the upwind scheme
(recipes BI \& BII) efficiently eliminates high frequency noise.  This
property is very helpful in keeping the code stable, but the solution
with low resolution can be very inaccurate. Therefore, higher
resolution and, thus, greater computing resources are needed (compared
to the centered scheme) to achieve a desired accuracy.

%========================================
\subsection{3D results in boosted frames}
%========================================

We now describe the boosted case with the boost speed $v=0.5$. In this
case the BH, initially located in the origin of the lab frame, is
boosted along the $x$-axis with the boost velocity ${\bf
v}=(0.5,0,0)$. We stop the evolution when the separation between the
center of BH and the boundary is less than $4M$ to avoid the BH being
too close to the boundary. Since the origin of the lab frame is set
such that the positive boundary of $x$-axis is at $x=28M$, the total
evolution time is $24M/0.5=48M$.

Fig.~\ref{fig5} shows the waveforms of the scalar field.  The boosted
1D results are derived from the non-boosted (comoving) 1D result with
a Lorentz transformation (see Appendix~\ref{transn}).  We observe the
scalar waveforms at three check points, located at ($14M$,$0$,$0$),
($0$,$14M$,$0$), and ($-14M$,$0$,$0$).  In the first column of
Fig.~\ref{fig5}, there are two peaks in each waveform observed at
($14M$,$0$,$0$).  The first peak is the incoming wave packet, and the
second peak is the outgoing wave packet.  This is more obvious in the
2D snapshots in Fig.~\ref{fig6} (see also Fig.~\ref{fig1}).  The
waveforms are masked at $t=26.3M - 29.7M$ in the first column of
Fig.~\ref{fig5} because the excision zone passes over the check point
during that time.  The waveforms in the first column have larger
amplitude changes from low resolution to high resolution.  The
situation is especially obvious for recipe BII (using the upwind
scheme) in which the peak of $\Phi$ varies from $0.32$ at low
resolution to $0.55$ at high resolution.  We restricted the high
resolution run to a computational domain
($-8M<x<24M$,$-16M<y<16M$,$0<z<16M$) in order to save computational
resources.  Accordingly, no high resolution results are available for
the check point $(-14M,0,0)$.  Fig.~\ref{fig5} shows that the 3D
results of AII and BII still converge to the 1D solution although the
rate of convergence for both recipes is somewhat worse than in the
non-boosted case.

Recipe BI produces an instability for boosted BHs.  We believe that
this instability arises as a result of combining the simple inner
boundary condition of recipe BI with the treatment of gridpoints that
newly emerge from inside the excision zone.  The error expands outward
as the BH moves.  Physically, nothing can emerge from the BH, but
numerically noise can propagate outward through the event
horizon. Since the third-order extrapolation (\ref{3rdex}) condition
is equivalent to using a one-sided scheme at the excision boundary, it
propagates any numerical errors {\it inwards}.  We believe that
this extrapolation produces the improved behavior seen in recipe BII.
Furthermore, Eq.~(\ref{3rdex}) is viable in all cases (non-boosted and
boosted) as long as the excision zone is inside the event horizon.

AI is stable in the $v=0.2$ case but produces instability in $v=0.5$
case.  We believe that this instability is again caused by the
treatment of the grid points emerging from inside the excision zone,
where numerical errors are allowed to propagate outwards by the
centered scheme.  The instability can be avoided by using the upwind
scheme, which propagate the numerical errors inwards (e.g., recipe
AII).

Recipe BII is stable in all the cases.  Fig.~\ref{fig6} shows the
stable evolution of the scalar field with recipe BII. The excision
zone can be seen in the snapshots at $t=12M$, $t=16M$ and $t=24M$
where the residual wave amplitude is seen to hover about the excision
zones centered at ($x$,$y$)$=$($6M$,$0$), ($8M$,$0$), and ($12M$,$0$)
before disappearing entirely at later time inside the masked region.
Recipe AII is also stable in all the cases, although the numerical
errors in AII are bigger than in BII.  We have also confirmed that
recipes AII and BII are stable for boosts in off-axis directions by
studying a boost of $v=0.5$ in the ${\bf v}=(0.4,0.3,0)$ direction.

%========================================
\section{conclusion}
%========================================
\label{conc}
%========================================

We have compared four different numerical recipes for numerically
evolving a scalar field in a Kerr-Schild BH spacetime.  We have
integrated the equations both in comoving and lab frames. We study
the stability and convergence with the goal of identifying simple
promising techniques for more general evolution problems.  We used
singularity excision and have experimented with the boundary
conditions for the excision zone.

Our results show that, in general, an upwind scheme is superior to a
centered scheme for maintaining stability. This is consistent with the
results of \cite{alcm2}.  We find that higher resolution is needed for
an upwind scheme than for a centered one to achieve a desired accuracy
due to the diffusive character of an upwind scheme.

As summarized in Table~\ref{reps}, two of the four recipes are stable
in all testbed scenarios: recipe AII, in which the upwind scheme is
used on the shift (advection) term only inside the event horizon, and
recipe BII, in which the upwind scheme is used on the shift term
everywhere, as suggested by \cite{alcm2}. Both AII and BII use an
extrapolation boundary condition for treating the excision zone.
However, recipe BI, which follows the implementation of \cite{alcm2}
and simply copies time derivatives as opposed to extrapolating at the
boundary of the excision zone, fails in the boosted case.  The
instability is probably caused by the inner boundary condition and its
treatment of the grid points emerging from inside the excision zone.
We find that stability requires a higher-order extrapolation condition
to update the values of the grid points at the boundary of the
excision zone when the BH moves.

Our study may be useful for the numerical evolution of binary BHs. It
suggests that a simpler scheme (compared with the causal differencing
schemes) might be implemented to handle evolution and excision for a
moving BH.  Our scalar field model problem provides necessary criteria
for stability of a numerical recipe.  We are now gearing up to apply
these simple but stable numerical recipes to a full dynamical code
containing black holes.

%========================================
\acknowledgments
%========================================

It is a pleasure to thank M. Saijo and M. Duez for many helpful
discussions. This work was supported in part by NSF Grants PHY 99-02833
and PHY 00-90310,
and NASA Grants NAG 5-10781 and NAG 5-8418 at the University of Illinois at
Urbana-Champaign (UIUC).  Much of the calculation was performed at the
National Center for Supercomputing Applications at UIUC.
TWB gratefully acknowledges financial support through a Fortner Fellowship;
HJY acknowledges the support of the Academia Sinica, Taipei.

\appendix

%========================================
\section{Stability analysis}
%========================================
\label{staa}
%========================================

To analyze the 1D radial field equation (\ref{sfe2}) for stability,
we simplify it by retaining only the highest order derivative term
in $f$ in Eq.~(\ref{1dfe}), i.e.,
\begin{equation}\label{1dsim}
   \left\{
   \begin{array}{rcl}
      \Phi_{,t}    &=& \beta^r\Phi_{,r} + \pi,\\
      \pi_{,t}  &=& \beta^r\pi_{,r}  + a \Phi_{,rr},
   \end{array}
   \right.
\end{equation}
where $a=1/(1+2H)^2$.

%========================================
\subsection{The Centered Scheme}
%========================================

We use an iterated Crank-Nicholson method with one predictor step and 
two corrector steps to evolve Eq.~(\ref{1dsim}) (see~\cite{teus}).
To 
apply a von Neumann stability analysis we assume
\begin{equation} \label{ansatz}
   \begin{array}{rcl}
      \Phi^n_j&=&\hat{\Phi}\xi^ne^{ikj\Delta x},\\
      \pi^n_j&=&\hat{\pi}\xi^ne^{ikj\Delta x},
   \end{array}
\end{equation}
and adopt the following finite differencing for the predictor step
(``forward time centered space'') 
\begin{eqnarray}
   \Phi_{,t} &\rightarrow& {1\over\Delta t} (\Phi^{n+1}_j - \Phi^n_j),
      \nonumber \\
   \pi_{,t} &\rightarrow& {1\over\Delta t} (\pi^{n+1}_j-\pi^n_j),
      \nonumber \\
   \pi &\rightarrow&  \pi^n_j, \nonumber \\
   \pi_{,r} &\rightarrow&  {\pi^n_{j+1} - \pi^n_{j-1} \over2\Delta x}
      = {i\sin k\Delta x\over\Delta x} \pi^n_j, \label{cs1} \\
   \Phi_{,r} &\rightarrow& {\Phi^n_{j+1} - \Phi^n_{j-1} \over2\Delta x}
      = {i\sin k\Delta x\over\Delta x} \Phi^n_j, \nonumber \\
   \Phi_{,rr} &\rightarrow& {\Phi^n_{j+1} - 2\Phi^n_j + \Phi^n_{j-1}
      \over(\Delta x)^2}  \nonumber \\
   && = {2\over(\Delta x)^2} (\cos k\Delta x-1)\Phi^n_j.  \nonumber
\end{eqnarray}
Here $n$ labels the time level and $j$ labels the spatial grid point.
Substituting (\ref{cs1}) into (\ref{1dsim}) yields
\begin{equation}\label{veceq}
   {{\bf u}^{n+1}_j - {\bf u}^n_j \over\Delta t} = {\bf A}{\bf u}^n_j,
\end{equation}
where
\begin{equation}
   \begin{array}{l}
      {\bf u}^n_j = \left(
      \begin{array}{c}
         \Phi^n_j\\
         \pi^n_j
      \end{array}
      \right), \\
      \noalign{\vskip0.2cm}
      {\bf A} = \left(
      \begin{array}{cc}
         2i\beta^r P&1 \\
         2aQ&2i\beta^r P
      \end{array}
      \right), \\
      \noalign{\vskip0.2cm}
      \quad P=\displaystyle{\sin k\Delta x\over2\Delta x},
      \quad Q=\displaystyle{\cos k\Delta x-1\over(\Delta x)^2}.\\
   \end{array}
\end{equation}
To find the eigenvalues $\lambda$'s for the matrix ${\bf A}$, we require
\begin{equation}\label{detm}
   \det({\bf A}-\lambda{\bf I})=0,
\end{equation}
which gives
\begin{equation}
   ( \lambda -2i\beta^r P )^2 = 2 aQ,
\end{equation}
or
\begin{equation}\label{eigens}
   \lambda  = 2i{Z\over\Delta t},
\end{equation}
where
\begin{eqnarray}
   Z &=& {\Delta t\over \Delta x}\sin (k\Delta x/2)
         (\beta^r \cos (k\Delta x/2) \pm \sqrt{a} ) \nonumber \\
     &=& {\Delta t\over \Delta x}{\sin (k\Delta x/2)\over (1+2H)}
         \left[ 2H \cos (k\Delta x/2) \pm 1 \right].
\label{zzz}
\end{eqnarray}
Here $|Z|\le1$ for the value of the shift ($\beta^r<1$) and $a$
if $\Delta t/\Delta x\le 1$.

For the convenience of analysis, we would like to decouple
the equation set (\ref{veceq}) into two independent equations.
By using a suitable coordinate transformation $R$
we can diagonalize the matrix ${\bf A}$
\begin{equation}
   {\bf A}'=R{\bf A}R^{-1}=2i{Z\over\Delta t}{\bf I},
\end{equation}
so that Eq.~(\ref{veceq}) decouples into two independent
equations for the components of the vector ${\bf u}'=R{\bf u}$, 
\begin{equation}\label{decoupfe}
   {\bf u}'^{n+1}_j - {\bf u}'^n_j =2iZ{\bf u}'^n_j,
\end{equation}
where $Z$ takes two different values (i.e., different choices of sign 
in Eq.~(\ref{zzz})) for the two different components of ${\bf u}'$.

Now we can use the procedure described in \cite{teus} to derive the
spectral radius (amplification factor) $\xi$.  The first iteration of
the iterated Crank-Nicholson method starts by calculating an
intermediate variable ${}^{(1)}\tilde{\bf u}$ using
Eq.~(\ref{decoupfe})
\begin{equation}
  {}^{(1)}\tilde{\bf u}^{n+1}_j - {\bf u}^n_j =2iZ{\bf u}^n_j,
\end{equation}
where we have dropped the prime for convenience.  Averaging to 
an intermediate time level $n+1/2$ yields
\begin{equation}
   {}^{(1)}\tilde{\bf u}^{n+1/2}_j
    ={1\over2}({}^{(1)}\tilde{\bf u}^{n+1}_j+{\bf u}^n_j)
    =(1+iZ){\bf u}^n_j.
\end{equation}
This value is now used in the first corrector step, in which 
${}^{(1)}\tilde{\bf u}^{n+1/2}$ is used on the right hand side of
Eq.~(\ref{decoupfe})
\begin{eqnarray}
   {}^{(2)}\tilde{\bf u}^{n+1}_j - {\bf u}^n_j &=&
   2iZ{}^{(1)}\tilde{\bf u}^{n+1/2}_j = 2iZ(1+iZ){\bf u}^n_j.\\
   {}^{(2)}\tilde{\bf u}^{n+1/2}_j &=& {1\over2}({}^{(2)}\tilde{\bf
   u}^{n+1}_j+{\bf u}^n_j) \nonumber \\ &=& (1+iZ-Z^2){\bf u}^n_j.
\end{eqnarray}
The second (and final) corrector step now uses ${}^{(2)}\tilde{\bf
u}^{n+1/2}$ on the right-hand side of Eq.(\ref{decoupfe})
\begin{eqnarray}
   {\bf u}^{n+1}_j - {\bf u}^n_j
    &=& 2iZ{}^{(2)}\tilde{\bf u}^{n+1/2}_j \nonumber \\
    &=& 2iZ(1+iZ-Z^2){\bf u}^n_j.
\end{eqnarray}
Inserting~(\ref{ansatz}) we finally find
\begin{equation}
   \xi = 1+2iZ-2Z^2-2iZ^3.
\end{equation}
Since $|Z|\le1$ provided $\Delta t/\Delta x\le 1$,
we find $|\xi|\le1$ so that the centered differencing scheme is stable.

%========================================
\subsection{The Upwind Scheme}
%========================================

Here we apply a first-order upwind scheme only to the shift term.
The finite differencing (\ref{cs1}) won't change except for the terms
$\Phi_{,r}$ and $\pi_{,r}$, where
\begin{eqnarray}
   \Phi_{,r} &\rightarrow& {\Phi^n_{j+1} - \Phi^n_j \over\Delta x}
    = S \Phi^n_j, \nonumber \\
   \pi_{,r} &\rightarrow&  {\pi^n_{j+1} - \pi^n_j \over\Delta x}
    = S \pi^n_j,
\end{eqnarray}
where $S=\displaystyle{e^{ik\Delta x}-1\over\Delta x}$.
Substituting into (\ref{1dsim}) now yields
\begin{equation}\label{vecos}
   {{\bf u}^{n+1}_j - {\bf u}^n_j \over\Delta t} = {\bf B}{\bf u}^n_j,
\end{equation}
where
\begin{equation}
   {\bf B}=\left(
   \begin{array}{cc}
      \beta^r S & 1       \\
      2aQ        & \beta^r S
   \end{array}
   \right).
\end{equation}
With the same argument in the last section Eq.~(\ref{vecos})
can be rewritten as
\begin{equation}
   {\bf u}^{n+1}_j - {\bf u}^n_j = 2W{\bf u}^n_j,
\end{equation}
where
\begin{eqnarray}
   W &=& iZ-Y,\nonumber \\
   Y &=& {\Delta t\over\Delta x}\beta^r\sin^2(k\Delta x/2)
      =  {\Delta t\over\Delta x}{2H\sin^2(k\Delta x/2)\over (1+2H)},\\
   Z &=& \mbox{rhs of Eq.(\ref{zzz}).}\nonumber
\end{eqnarray}
Following the same procedure as in last section
one finds the spectral radius 
\begin{equation}\label{1stsr}
   \xi = 1+2W+2W^2+2W^3,
\end{equation}
which again satisfies $|\xi|\le1$ provided $\Delta t/\Delta x\le 1$.
The first-order one-sided scheme is therefore stable as well.  For our
second-order one-sided scheme, we must change the finite differencing
in (\ref{cs1}) to
\begin{eqnarray}
   \Phi_r &\rightarrow& {-\Phi^n_{j+2} + 4\Phi^n_{j+1} - 3\Phi^n_j
      \over 2\Delta x},\nonumber\\
   \pi_r &\rightarrow& {-\pi^n_{j+2} + 4\pi^n_{j+1}
- 3\pi^n_j \over 2\Delta x}.
\end{eqnarray}
Analytic analysis becomes complicated in this case, so we rely on
empirical results, which again exhibit stability.

%========================================
\section{Lorentz transformation of a scalar wave}
%========================================
\label{transn}
%========================================

In order to produce the initial data for the boosted cases as well as
the ``exact'' boosted waveforms to compare with the 3D numerical
results, we need to know the transformation of the scalar wave
solution in the 1D static BH background to the solution as viewed in a
frame in which the BH is boosted.

Assume a boost velocity ${\bf v}=(v_1,v_2,0)$.
The Lorentz transformation between the comoving frame $X^\prime$ and the
lab frame $X$ is $ x'^\beta=\Lambda^\beta{}_\alpha x^\alpha$, i.e.,
\begin{equation}\label{lf2}
\begin{array}{rcl}
t^\prime&=&\gamma t-\gamma v_1x-\gamma v_2y,\\
x^\prime&=&x+\displaystyle{\gamma^2v_1\over\gamma+1}(v_1x+v_2y)
-\gamma v_1t,\\
y^\prime&=&y+\displaystyle{\gamma^2v_2\over\gamma+1}(v_1x+v_2y)
-\gamma v_2t,\\
z^\prime&=&z,
\end{array}
\end{equation}
where $\gamma=1/\sqrt{1-v^2}$ and $v^2=v^2_1+v^2_2$.
Since $\phi$ is a scalar, it transforms according to 
$\Phi'(x'^\beta)=\Phi(x^\alpha)$.
 
For the initial data, we also need to know $\pi(t=0,{\bf x})$ using 
\begin{equation}
   {\partial\Phi\over\partial x^\mu}=\Lambda^\nu{}_\mu
   {\partial\Phi\over\partial x'^\nu},
\end{equation}
We obtain $\pi$ from its definition $\pi \equiv \Phi_{,t} -
\beta^i\Phi_{,i}$.  By using these formula we can derive the initial
data ($t=0$) and the 1D waveform ($t>0$) in the all boosted cases.

%========================================


\begin{thebibliography}{99}
\bibitem{fora}
For a boosted BH along the $x$-axis viewed in the lab frame,
the apparent horizon is located on the spheroidal boundary given by
$r^2=\gamma^2(x-vt)^2+y^2+z^2=(2M)^2$, see Eq.~(\ref{ksf3}).
\end{thebibliography}
\end{document}